# Do We Really Need Specialization? Evaluating Generalist Text Embeddings for Zero-Shot Recommendation and Search

MATTEO ATTIMONELLI, Politecnico Di Bari, Italy and Sapienza University of Rome, Italy
ALESSANDRO DE BELLIS, Politecnico Di Bari, Italy
CLAUDIO POMO, Politecnico Di Bari, Italy
DIETMAR JANNACH, University of Klagenfurt, Austria
EUGENIO DI SCIASCIO, Politecnico Di Bari, Italy
TOMMASO DI NOIA, Politecnico Di Bari, Italy

Pre-trained language models (PLMs) are widely used to derive semantic representations from item metadata in recommendation and search. In sequential recommendation, PLMs enhance ID-based embeddings through textual metadata, while in product search, they align item characteristics with user intent. Recent studies suggest task and domain-specific fine-tuning are needed to improve representational power. This paper challenges this assumption, showing that *Generalist Text Embedding Models (GTEs)*, pre-trained on large-scale corpora, can guarantee strong zero-shot performance without specialized adaptation. Our experiments demonstrate that GTEs outperform traditional and fine-tuned models in both sequential recommendation and product search. We attribute this to a superior representational power, as they distribute features more evenly across the embedding space. Finally, we show that compressing embedding dimensions by focusing on the most informative directions (e.g., via PCA) effectively reduces noise and improves the performance of specialized models. To ensure reproducibility, we provide our repository at https://split.to/gte4ps.

CCS Concepts: • **Do Not Use This Code** → **Generate the Correct Terms for Your Paper**; *Generate the Correct Terms for Your Paper*; Generate the Correct Terms for Your Paper; Generate the Correct Terms for Your Paper.

Additional Key Words and Phrases: Sequential Recommendation, Product Search, Generalist Text Embedding Models

## 1 Introduction

Recommendation and Search systems are essential for helping users navigate large information spaces, from retrieving relevant documents to providing personalized product suggestions. Traditional methods often rely on ID-based embeddings or keyword-based matching, which may struggle to capture the nuanced semantics of user intent and item content, especially in cold-start scenarios. Recent advances in Natural Language Processing, especially the rise of pretrained language models (PLMs), have created new opportunities for tackling these challenges. PLMs like BERT [5] and T5 [28] move beyond shallow matching by capturing rich representations of queries, items, and users. In recommendation, particularly in sequential settings, they enhance item embeddings by modeling content and user behavior dynamics [11, 36]; in search, they improve query-item alignment. Despite their benefits, PLMs are not directly optimized for recommendation and search tasks, often limiting generalization [10]. Recent work has addressed this through task-specific fine-tuning. One notable example is BLAIR [10], a RoBERTa [20]-based embedding model, fine-tuned on large-scale user reviews and item metadata to better align query and item embeddings, improving efficacy in sequential recommendation and product search. While domain and task adaptation methods such as BLAIR highlight their benefits of specialized models, recent advances in text embedding techniques have led to the emergence of *Generalist Text*

Authors' Contact Information: Matteo Attimonelli, matteo.attimonelli@poliba.it, Politecnico Di Bari, Bari, Italy and Sapienza University of Rome, Rome, Italy; Alessandro De Bellis, alessandro.debellis@poliba.it, Politecnico Di Bari, Bari, Italy; Claudio Pomo, claudio.pomo@poliba.it, Politecnico Di Bari, Bari, Italy; Dietmar Jannach, dietmar.jannach@aau.at, University of Klagenfurt, Klagenfurt, Austria; Eugenio Di Sciascio, eugenio.disciascio@poliba.it, Politecnico Di Bari, Bari, Italy; Tommaso Di Noia, tommaso.dinoia@poliba.it, Politecnico Di Bari, Bari, Italy.

*Embedding Models (GTEs)* [17, 19]. Trained on diverse corpora and tasks, GTEs learn broad, transferable semantics, offering a training-free alternative to fine-tuning at scale. Thus, they may offer a strong alternative as embedding models for recommendation and search, achieving competitive performance without specific adaptation.

This motivates the main research question in this paper: **How well do GTEs perform in recommendation and search tasks in a zero-shot setting, compared to traditional and fine-tuned models?** To address this question, we evaluate state-of-the-art GTEs on two key tasks: *sequential recommendation* and *product search*, focusing on models included in the *Massive Text Embedding Benchmark (MTEB)* [24]. Our results show that GTEs outperform both traditional and fine-tuned models, with recent open-source models (e.g., NVEmbed-v2 [17] and GTE-Qwen2 [19]) even surpassing popular closed-source alternatives such as OpenAI's.

Moreover, our study takes a first step towards addressing a largely open question: *What underlying characteristics of textual embeddings influence their effectiveness for recommendation and search?* We focus on factors related to embedding geometry and dimensionality, aiming to uncover the key characteristics that govern embedding effectiveness. In particular, we analyze *space utilization*, examining how effectively models distribute variance across embedding dimensions, addressing concerns about *dimensional collapse* [6]. We use PCA to estimate the *effective dimensionality*, providing a quantitative measure of the phenomenon. This analysis reveals that GTEs tend to achieve more uniform space utilization. Building on this, we demonstrate that PCA itself provides a pathway for effective compression: retaining only the most informative components substantially reduces GTE dimensionality without sacrificing accuracy, thus improving scalability. We further find that this compression also benefits fine-tuned models, removing noisy dimensions and reducing the performance gap compared to GTEs.

The rest of the paper is organized as follows: Section 2 provides an overview of related work, Section 3 describes the methodology and tasks, Section 4 presents the experimental setup and results, and Section 5 concludes the paper with a discussion of future directions.

## 2 Related Work

*Generalist Text Embedding Models.* GTEs [19] are text embedding models trained on large-scale corpora and diverse tasks to produce rich, transferable representations for a variety of downstream applications. Several encoder-based GTEs build on the transformer encoder [5] architecture. For instance, mGTE [39] extends BERT [5] with RoPE [33] and unpadding [27] for long-context support. INSTRUCTOR [32] enhances GTR [26] via instruction tuning for task-aware embeddings, while Sentence-T5 [25] derives embeddings from the T5 [28] encoder block. Decoder-only LLMs have also been adapted for embedding tasks. GTE-Qwen2 [19] repurposes Qwen2 [35] via multi-stage training to generate universal embeddings. Jasper [37] applies Matrioshka Representation Learning [16] to distill embeddings from a larger teacher model. NVEmbed-v2 [17], based on Mistral-7B [13], removes causal masking and introduces latent attention for improved token-aware pooling. KALM [12], being fine-tuned on a small model (i.e., Qwen2-0.5B), shows that even compact models can yield high-quality embeddings when trained on large-scale, multi-task datasets.

*Text Encoders for Sequential Recommendation and Product Search.* Sequential recommendation predicts a user's next interaction based on past behavior. Transformer-based models (e.g., SASRec [14] and BERT4Rec [34]) address the limitations of recurrent models such as GRU4Rec [9], employing self-attention for long-range dependencies and parallelization. Yet, many approaches rely solely on item IDs, ignoring side information like metadata and images. UniS-Rec [11] demonstrates the effectiveness of incorporating multimodal signals, combining ID and semantic features [36]. Unlike recommendation, product search requires explicit query-item matching. Traditional methods (e.g., BM25) have been supplanted by deep models that learn dense query and product embeddings. Transformer-based approaches like BERT [5] and ColBERT [15] enhance semantic matching but face limitations due to a certain scarcity of public datasets [22, 23]. To address this, BLAIR [10] introduces a contrastive learning-based text encoder using reviews as pseudo-queries. Still, most text embedding solutions in recommendation and search rely on closed-source models [3, 8] or domain-specific fine-tuning (e.g., BLAIR) leaving GTEs largely unexplored.

*Embedding Quality Evaluation.* Embedding models are commonly evaluated implicitly through downstream task performance. MTEB [24] provides a comprehensive suite for assessing the performance of GTEs across such tasks. Beyond downstream performance, it is also crucial to investigate the intrinsic properties of embedding models, as these can significantly influence their effectiveness and suitability in practical applications. Factors such as embedding dimension and model capacity affect not only accuracy but also the scalability and sustainability of retrieval systems [16]. Another key aspect is space utilization: a frequent issue in contextual pre-trained language models is *dimensional collapse* [6, 7], where embeddings lie in low-rank subspaces, limiting expressiveness and degrading performance [21] in tasks reliant on distance metrics [25]. Consequently, post-processing methods [18, 38] have been proposed to improve embedding utility. In this work, we investigate whether GTEs inherently possess properties that enable efficient space utilization, potentially eliminating the need for ad-hoc adaptation.

## 3 Background and Methodology

This section outlines our methodology and formalizes two core tasks: sequential recommendation and product search.

### 3.1 Task Overview

*3.1.1 Sequential Recommendation (SR).* Sequential recommendation focuses on modeling a user's evolving interests based on their interaction history, with the goal of predicting the next item of interest. Let a user session be represented as a sequence $\mathcal{S}_t = [i_1, i_2, \ldots, i_t]$, $i_j \in \mathcal{I}$, where $\mathcal{I}$ is the set of all items, and $i_j$ denotes the item interacted with at position $j$. The recommendation task can be framed as learning a function $f_{\text{seq}}(\mathcal{S}_t) \to i_{t+1}$. Recent approaches employ neural sequence models to encode the interaction sequence. Each item $i_j$ is mapped to an embedding $\mathbf{e}_{i_j} \in \mathbb{R}^d$ using an item encoder $\mathbf{e}_i$, and a sequence encoder $\phi_{\text{seq}}$ transforms the embedded sequence into a contextual representation $\mathbf{h}_t = \phi_{\text{seq}}([\mathbf{e}_{i_1}, \mathbf{e}_{i_2}, \ldots, \mathbf{e}_{i_t}])$, where $\mathbf{h}_t \in \mathbb{R}^d$ captures the user's dynamic preference at time $t$. This representation $\mathbf{h}_t$ is then used to score or rank the candidate items via a similarity measure, $\text{Sim}(\cdot, \cdot)$, as $f_{\text{seq}}(\mathcal{S}_t) = \arg\max_{i \in \mathcal{I}} \text{Sim}(\mathbf{h}_t, \phi_I(i))$.

*3.1.2 Product Search (PS).* Product search aims to retrieve the most relevant items from a large catalog $\mathcal{I} = \{i_1, i_2, \ldots, i_N\}$ in response to a user query $q \in Q$ (the space of possible queries), often expressed in natural language. Recent approaches significantly improve upon traditional lexical matching (e.g., BM25) by leveraging PLMs to map queries and items into a shared high-dimensional embedding space. A pre-trained text embedding model generates dense vector representations $\mathbf{e}_q$ for queries and $\mathbf{e}_i$ for items, typically using their textual attributes (like titles, descriptions, or metadata). The core task is to learn a scoring function $f(q, i) \to \mathbb{R}$ that reflects the relevance of item $i$ to query $q$. This score is commonly computed as the similarity between their respective embeddings: $f(q, i) = \text{Sim}(\mathbf{e}_q, \mathbf{e}_i)$ where $\text{Sim}(\cdot, \cdot)$ often represents cosine similarity or dot product. Items are then ranked according to this relevance score.

## 3.2 Embedding Space Utilization

Text embedding models, which map text to dense representations, often suffer from *dimensional collapse* [6], where variance concentrates in a small subset of dimensions. We use *effective dimensionality* [4] to quantify this effect. Let $\mathbf{E}_\mathcal{I}^c \in \mathbb{R}^{N \times d}$ denote the set of mean-centered item embeddings, extracted from a pre-trained encoder using item metadata. Applying PCA yields singular values $\{\sigma_i\}_{i=0}^{d-1}$ for the covariance matrix of $\mathbf{E}_\mathcal{I}^c$. The **explained variance ratio** for the top $k$ components is defined as $r_k = \sum_{i=0}^{k-1} \sigma_i / \sum_{i=0}^{d-1} \sigma_i$. The $\varepsilon$-**effective dimension** is then $d(\varepsilon) \coloneqq \arg\min_k \; r_k \geq \varepsilon$, representing the minimum number of components needed to retain at least an $\varepsilon$ fraction of the total variance. Following prior work [4], we robustly estimate principal components by applying PCA to a random subset $\hat{\mathbf{E}}_\mathcal{I}^c \in \mathbb{R}^{n \times d}$, where $n \gg d$ [30]. Projecting query embeddings onto the resulting subspace ensures representational consistency, reduces dimensionality, and preserves the most discriminative features. This analysis (i) quantifies **space utilization** [30], which impacts downstream performance [21], and (ii) establishes a **lower bound for compression** critical for scaling large GTEs while maintaining informativeness.

## 4 Experiments and Discussion

We evaluate different GTEs on sequential recommendation (SR) and product search (PS), as defined in Section 3.1, and analyze the properties of their embeddings in relation to performance. Our experiments address two core research questions:

**RQ1:** How well do GTEs perform in recommendation and search tasks in a zero-shot setting, compared to traditional and fine-tuned models?

**RQ2:** What underlying characteristics of textual embeddings influence their effectiveness for recommendation and search?

### 4.1 Experimental Setup

*4.1.1 Datasets.* To ensure fair comparison with prior work, we adopt the experimental configuration of Hou et al. [10] for both SR and PS. We use the Amazon Reviews 2023 dataset [10], accessed via the HuggingFace[1] `datasets` API, following the original data splits and preprocessing. Item metadata is constructed by concatenating the *title*, *features*, *categories*, and *description* fields.

For SR, we focus on the *All Beauty* (Beauty), *Video Games* (Games), and *Baby Products* (Baby) categories, with user-item interactions ranging from 104,766 to 3,583,323, reflecting varying data sparsity. For PS, we use the ESCI [29] and Amazon-C4 [10] datasets. ESCI provides real-world `<query,item>` pairs with graded relevance labels. Following Hou et al. [10], we select only *exact* matches from the English *ESCI-small* split for higher-quality supervision, containing 27,643 queries and 1,367,729 items. Amazon-C4 comprises 21,223 queries and 1,058,417 items, with synthetic queries generated from 5-star reviews using ChatGPT [2]. Queries in both datasets are linked to Amazon Reviews 2023 via product IDs. We report results on two representative categories: *Office Products* (Office) and *Sports and Outdoors* (Sports). Following the evaluation protocol of Hou et al. [10], each `<query,item>` pair is ranked against a pool of 50 randomly sampled items for each domain.

*4.1.2 Models.* For SR, we consider the baseline models GRU4Rec [9], SASRec [14], and UniSRec [11]. Both GRU4Rec and SASRec are designed to operate with item ID embeddings. To incorporate textual information, we follow the

---
[1] https://huggingface.co/datasets/McAuley-Lab/Amazon-Reviews-2023

Table 1. Sequential recommendation performance: Recall@k ($R_k$) and nDCG@k ($N_k$). Results are reported as percentages. Best scores are bolded, second-best underlined. ID stands for ID-based, T stands for Text-based.

| Model | Beauty | | | | Games | | | | Baby | | | |
|---|---|---|---|---|---|---|---|---|---|---|---|---|
| | $R_{10}$ | $N_{10}$ | $R_{50}$ | $N_{50}$ | $R_{10}$ | $N_{10}$ | $R_{50}$ | $N_{50}$ | $R_{10}$ | $N_{10}$ | $R_{50}$ | $N_{50}$ |
| **GRU4Rec$^{ID}$** | 0.31 | 0.15 | 1.11 | 0.32 | 2.11 | 1.14 | 5.28 | 1.82 | 1.10 | 0.56 | 3.55 | 1.09 |
| **SASRec$^{ID}$** | 0.31 | 0.18 | 0.84 | 0.30 | 2.32 | 1.15 | 5.55 | 1.85 | 1.23 | 0.58 | 3.69 | 1.11 |
| **GRU4Rec$^{T}$** | | | | | | | | | | | | |
| BLaIR$_B$ | 0.63 | 0.30 | 1.67 | 0.53 | 2.16 | 1.14 | 5.43 | 1.84 | 1.29 | 0.65 | 4.05 | 1.24 |
| BLaIR$_L$ | 0.84 | 0.42 | 2.09 | 0.69 | 2.35 | 1.29 | 5.90 | 2.06 | 1.26 | 0.64 | 3.96 | 1.22 |
| t-emb-3 | 0.57 | 0.27 | 1.78 | 0.55 | 3.01 | 1.64 | 7.39 | <u>2.59</u> | 0.44 | 0.21 | 1.67 | 0.47 |
| NVEmb$^†$ | 0.75 | 0.36 | 1.74 | 0.57 | 2.98★ | <u>1.65★</u> | 7.27★ | 2.57★ | <u>1.63★</u> | <u>0.84★</u> | **4.92★** | <u>1.54★</u> |
| KALM | 0.69 | 0.36 | 1.76 | 0.59 | 2.67 | 1.46 | 6.52 | 2.30 | 1.45 | 0.73 | 4.51 | 1.38 |
| **SASRec$^{T}$** | | | | | | | | | | | | |
| BLaIR$_B$ | 0.69 | 0.35 | 2.16 | 0.66 | 1.90 | 1.01 | 5.01 | 1.68 | 1.24 | 0.63 | 3.84 | 1.19 |
| BLaIR$_L$ | 1.21 | 0.50 | 2.55★ | 0.77 | 2.18 | 1.17 | 5.36 | 1.86 | 1.33 | 0.68 | 4.04 | 1.26 |
| t-emb-3 | 0.84 | 0.42 | 2.20 | 0.71 | 2.77 | 1.51 | 6.92 | 2.41 | 0.40 | 0.20 | 1.43 | 0.42 |
| NVEmb$^†$ | 0.77 | 0.43 | 2.05 | 0.70 | 2.78★ | 1.54★ | 6.87★ | 2.42★ | 1.50★ | 0.75★ | 4.51★ | 1.40★ |
| KALM | 0.88 | 0.45 | 1.80 | 0.65 | 2.40 | 1.30 | 5.83 | 2.04 | 1.44 | 0.73 | 4.38 | 1.35 |
| **UniSRec$^{T}$** | | | | | | | | | | | | |
| BLaIR$_B$ | 2.60 | 1.39 | 4.69 | 1.85 | 2.47 | 1.30 | 6.08 | 2.07 | 1.52 | 0.77 | 4.34 | 1.37 |
| BLaIR$_L$ | 2.51 | 1.38 | 4.54 | 1.83 | 2.53 | 1.37 | 6.24 | 2.17 | 1.53 | 0.78 | 4.36 | 1.39 |
| t-emb-3 | **3.60★** | **1.86★** | **5.73★** | **2.31★** | <u>3.05</u> | 1.64 | <u>7.46</u> | 2.59 | 0.44 | 0.24 | 1.57 | 0.48 |
| NVEmb$^†$ | 2.93 | 1.58 | 4.77 | 2.00 | **3.11★** | **1.65★** | **7.50★** | **2.59★** | **1.72★** | **0.89★** | <u>4.87★</u> | **1.57★** |
| KALM | <u>3.45</u> | <u>1.84</u> | <u>5.55</u> | <u>2.29</u> | 2.75 | 1.47 | 6.68 | 2.32 | 1.55 | 0.78 | 4.39 | 1.39 |

($^†$) NVEmb is adopted as abbreviation of NVEmbed-v2.
($^★$) Statistically significant improvement (p < 0.001) between the best GTE model and the best BLaIR-based model.

strategy proposed by Hou et al. [10], using item descriptions as textual inputs. This results in two variants for each model: an **ID**-based version using standard embeddings and a **T**ext-based version with text-derived representations. For the text-based variants, we test several pre-trained text embedding models: BLaIR [10] (base and large), OpenAI's `text-embedding-3-large`[2] (t-emb-3) and two open-source GTEs: KALM [25] and NVEmbed-v2 [17]. KALM matches BLaIR in parameter count and embedding size for fair comparison, while NVEmbed-v2 offers a more powerful GTE with higher dimensionality and a larger parameter footprint.

For PS, we expand our evaluation to include a range of text encoders: RoBERTa$_{BASE}$ [20], ColBERTv2 [31], mGTE [39], INSTRUCTOR$_{XL}$ [32], Sentence-T5$_{XXL}$ [25], Jasper [37], and GTE-Qwen2 [19]. These models represent a mix of encoder- and decoder-based transformers, covering various parameter sizes and embedding dimensions for a comprehensive comparison.

*4.1.3 Evaluation Protocol.* For SR, we employ the RecBole [40, 41] framework, adopting the optimized hyperparameter settings from [10], as they are already well-tuned for the Amazon Reviews 2023 dataset and ensure fair comparison. Early stopping is applied based on nDCG@10 with a patience of 10. For PS, models are implemented using Hugging Face Transformers, with a maximum input sequence length of 512. Retrieval is performed using cosine similarity, which empirically outperforms Euclidean and dot product measures. The metrics employed are Recall and nDCG. We analyze embedding space utilization by computing the $\varepsilon$-effective dimension using PCA on a random sample of 100,000 item embeddings [30].

Table 2. Comparison of text encoders for product search (nDCG@100) on the two datasets (results are percentages). "All" averages scores for all domains. We also report two domains, "Office" and "Sports".

|  | Model | #Par | #Dim | ESCI All | ESCI Office | ESCI Sports | Amazon-C4 All | Amazon-C4 Office | Amazon-C4 Sports |
|---|---|---|---|---|---|---|---|---|---|
|  | BM25 | — | — | 1.10 | 1.35 | 1.57 | 0.00 | 0.00 | 0.00 |
|  | RoBERTa$_B$ | 123M | 768 | 0.08 | 0.03 | 0.25 | 0.25 | 0.33 | 0.37 |
|  | ColBERTv2 | 110M | 768 | 1.71 | 1.80 | 1.44 | 0.88 | 1.60 | 0.51 |
| *Encoders* | BLaIR$_B$ | 123M | 768 | 11.76 | 12.56 | 14.26 | 14.90 | 18.02 | 22.24 |
|  | BLaIR$_L$ | 354M | 1024 | 12.12 | 12.81 | 13.89 | 17.18 | 20.07 | 25.68 |
|  | mGTE | 305M | 768 | 22.17 | 22.32 | 24.12 | 13.11 | 15.33 | 17.38 |
|  | INSTRUCTOR$_{XL}$ | 1.5B | 768 | 20.53 | 20.13 | 21.32 | 12.22 | 15.79 | 16.38 |
|  | Sentence-T5$_{XXL}$ | 11B | 768 | 21.75 | 22.13 | 22.98 | 13.88 | 17.77 | 19.22 |
| *Decoders* | KALM | 0.5B | 896 | 23.12 | 23.76 | 24.39 | 15.61 | 18.55 | 22.01 |
|  | Jasper | 2B | 1024 | 26.62 | 26.07 | 27.11 | 17.10 | 20.23 | 23.16 |
|  | GTE-Qwen2 | 7B | 3584 | **28.06**★ | **28.09**★ | **29.09**★ | 17.92 | **21.41**★ | 24.03 |
|  | NVEmbed-v2 | 11B | 4096 | 27.59 | 27.24 | 28.45 | **19.02**★ | 21.32 | **25.91**★ |
|  | `t-emb-3` | — | 3072 | 25.76 | 24.67 | 26.96 | 17.53 | 20.86 | 25.23 |

(★) Statistically significant improvement (p < 0.001) between the best GTE model and the best BLaIR-based model.

### 4.2 Model Comparison (RQ1)

Table 1 presents the results for the SR task. Textual embeddings consistently outperform ID-based models, highlighting the importance of semantic item information. Among the embedding models, NVEmbed-v2 and KALM, both of the GTE family, deliver the most substantial gains across all architectures. NVEmbed-v2 achieves the best results on the Games and Baby categories, while KALM slightly outperforms NVEmbed-v2 on Beauty. Such improvements offered by leading GTE models are statistically significant compared to the best BLaIR baseline for most text-based architectures and datasets. The significance was evaluated by comparing the best GTE with the leading BLaIR variant for each model-dataset pair, based on nDCG@10. Furthermore, GTEs surpass the closed-source model `t-emb-3` on two datasets. These results showcase the benefits of recent, high-capacity embedding methods that do not depend on domain-specific knowledge.

Table 2 shows the PS results on the ESCI and Amazon-C4 datasets. GTE-based models consistently surpass traditional models (RoBERTa and ColBERTv2) and BLaIR. Notably, top-performing GTEs achieve statistically significant gains over the best-performing BLaIR baseline, validating the superiority of the GTE approach. Scores are generally higher on ESCI due to its simpler queries: GTE-Qwen2 improves from 0.1792 on Amazon-C4 to 0.2806 on ESCI, and Jasper from 0.1710 to 0.2662. An exception is BLaIR, which performs well on Amazon-C4 but drops sharply on ESCI, suggesting that fine-tuning hinders generalization. RoBERTa performs worse across the board, reflecting its limitations in retrieval tasks. Among the top performers are high-capacity GTEs, with NVEmbed-v2 leading on Amazon-C4 and GTE-Qwen2 on ESCI.

Overall, our findings indicate that **GTEs achieve strong performance in both tasks without requiring task-specific fine-tuning, outperforming traditional and fine-tuned solutions, as well as closed-source models**.

### 4.3 Model Analysis (RQ2)

To understand the superior performance of GTEs, we analyze model capacity, embedding structure, and architectural design. Table 2 shows that **performance does not scale linearly with model capacity** (no. of parameters) **and**

---
[2]API version: 2024-02-01

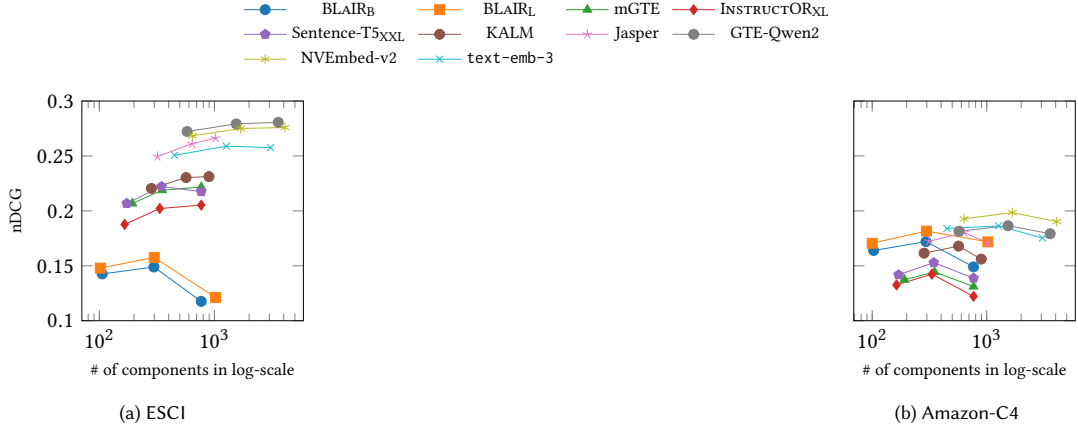

Fig. 1. nDCG scores for various models on the two PS datasets (a) ESCI and (b) Amazon-C4 versus the number of components (log-scale). Points shown correspond to component counts yielding 80%, 95%, and 100% explained variance.

**dimensionality**: KALM (0.5B) outperforms larger models such as INSTRUCTOR$_{XL}$ (1.5B) and Sentence-T5$_{XXL}$ (11B), while Jasper (1024 dimensions) outperforms t-emb-3 (3072 dimensions) on ESCI. In the SR task, KALM is found to outperform high-dimensional GTEs under specific model-dataset combinations (e.g., GRU4Rec-Beauty).

To assess space utilization, we compute the $\varepsilon$-effective-dimensionality (Section 3.2). Figure 1 shows how nDCG varies with the number of retained components in the PS task (points correspond to $\varepsilon$ values of 0.80, 0.95, and 1.00, left to right). An nDCG increase with fewer components suggests noisy directions in the feature space, while a drop indicates that information is uniformly distributed. We observe that for large GTEs (e.g., NVEmbed-v2 and GTE-Qwen2), the number of dimensions needed to preserve 80% of the total variance is close to the full dimensionality of BLaIR$_B$, suggesting more efficient use of representational space. BLAIR exhibits a significant drop in components with $\varepsilon = 0.80$, hinting at high anisotropy [4]. For GTEs, this phenomenon is less pronounced. While BLaIR benefits from PCA showing a marked inverted-U trend, many GTEs show a plateau or slight improvement at 95% variance: for large GTEs, moderate compression substantially reduces dimensionality without compromising performance. Conversely, on Amazon-C4, all models benefit from compression, resulting in a performance gain. This trend is particularly pronounced for BLaIR, indicating presence of noise in low-variance components.

Architecturally, we note that decoder-style models (Jasper, GTE-Qwen2, NVEmbed-v2) outperform encoder-based ones, with generative, autoregressive, or retrieval-oriented training yielding semantically rich, task-aligned representations [1, 17].

Overall, we show that **GTEs, particularly decoder-based ones, tend to spread useful information more evenly across embedding dimensions**, possibly explaining their strong performance. In contrast, we find that **model capacity and dimensionality often show little correlation with downstream performance**.

## 5 Conclusion and Future Work

This work investigates the zero-shot capabilities of GTEs in sequential recommendation and product search. Despite the potential advantages of domain-specific fine-tuning, we find that GTEs can outperform fine-tuned models (e.g., BLaIR), classical architectures (e.g., ColBERT), and closed-source solutions (e.g., OpenAI) without task-specific adaptation. We identify more uniform space utilization as a key factor for their success and show that emphasizing the most informative dimensions via PCA can boost the performance of fine-tuned models and effectively compress dimensionality for large GTEs. These results position GTEs as strong alternatives when fine-tuning is impractical. Future work may further enhance embedding properties such as isotropy and disentanglement, also explored in state-of-the-art models such as UniSRec.

Do We Really Need Specialization? Evaluating GTEs for Zero-Shot Recommendation and Search    9